\documentclass[12pt]{article}

\usepackage[utf8]{inputenc}
\usepackage[english]{babel}
\usepackage{textcomp}
\usepackage{amsfonts}
\usepackage{amsmath}
\usepackage{amssymb}
\usepackage{mathrsfs}
\usepackage[margin=2.2cm]{geometry}
\usepackage{graphics}
\usepackage{color}
\usepackage{epsfig}

\usepackage[numbers, sort&compress]{natbib}

\usepackage{slashed}

\usepackage{sidecap}
\usepackage{caption}
\usepackage{subcaption}

\usepackage[dvipsnames]{xcolor}

\begin{document}

\begin{center}
    {\bf \Large  Higgs alignment limits in the type-II 2HDM}\\[0.3cm]
    {\bf \Large and the MSSM with explicit CP-violation}
\end{center}

\begin{center}
{\large M.~N.~Dubinin}$^{1,2}$, 
{\large E.~Yu.~Fedotova}$^{1}$
\\
\vskip .7cm
{\footnotesize
$^{1}$ Skobeltsyn Institute of Nuclear Physics (SINP MSU), 
M.V. Lomonosov Moscow State University, Leninskie gory, GSP-1, 119991 Moscow, Russia\\[0.3cm]
$^2$ National University of Science and Technology MISIS,
Leninskiy Prospekt, 4, 119049, Moscow, Russia
\vskip .5cm
\begin{minipage}[l]{.9\textwidth}
\begin{center}
\textit{E-mail:}
\tt{dubinin@theory.sinp.msu.ru}, \tt{fedotova@theory.sinp.msu.ru}
\end{center}
\end{minipage}
}
\end{center}
\vskip 1cm
\begin{abstract}

For the general two-Higgs doublet model with Yukawa sector of type II (type II 2HDM), the Higgs alignment limit conditions are obtained for the neutral Higgs bosons with indefinite CP-parity $h_1, h_2$ or $h_3$, based on the symbolic results relating
the elements of the mixing matrix to the masses of the Higgs bosons and the mixing angles. 
The results are valid up to dimension-six operators in the decomposition of the effective Higgs potential. Within the framework of the obtained Higgs alignment conditions, the possibility of the existence of light scalars is discussed. 
Within the Minimal Supersymmetric Standard Model (MSSM) framework, four benchmark scenarios are proposed. It is shown that two of them predict phenomenologically distinguishable CP-violating interactions of the Higgs boson $h_3$ with up-fermions.

\end{abstract}

\vspace{0.2in}
\noindent
PACS: 12.60.Fr; 14.80.Da
\\
{\sc Keywords}: Higgs bosons with indefinite CP-parity; alignment limit; 2HDM; MSSM

\bigskip

\bigskip

\noindent
{\it To be published in International Journal of Modern Physics D as}
{doi:10.1142/S0218271825410020}

\newpage

\section{Introduction}	

The discovery of a Higgs boson at the CERN Large Hadron Collider (LHC) by ATLAS and CMS Collaborations in 2012 \cite{obs-125} confirms
the Brout-Englert-Higgs mechanism of electroweak symmetry breaking \cite{BEH-mech}. 
The measured properties of the observed Higgs boson are consistent with the expectations of the Standard
Model (SM) within the current experimental precision \cite{H_presic}. 
However, an admixture of CP-odd components to the Higgs boson mass state is still possible \cite{PDG-24}. 

Despite that the SM works extremely well, it has problems that cannot be solved within it. These include  neutrino oscillations, matter--antimatter asymmetry, strong CP problem, and the nature of dark matter, forcing us to consider
SM as an effective theory at low-energy.
At higher energies it may demonstrate new symmetries and include new fields, so the Higgs sector can be nonminimal and besides the SM Higgs doublet can contain additional Higgs multiplets.
In the simplest SM extension the Higgs sector includes two $SU(2)$ doublets (Two-Higgs Doublet Model, 2HDM) \cite{THDM} resulting in five physical Higgs states, two of them are charged $H^\pm$ and three states are neutral, in the CP-conserving limit (CPC) they are CP-even $h$ and $H$ states and a CP-odd state $A$; in a model with CP-violation (CPV) of the Higgs potential they are $h_1,h_2,h_3$ states with indefinite CP-parity \cite{echaja}. 

In any beyond-the-SM theory (BSM),
properties of the observed Higgs boson (within the precision of experiment) must satisfy the Higgs alignment limit conditions \cite{align}
\begin{equation}
m = 125 \text{ GeV}, \qquad 
g \equiv y_{\rm BSM}/y_{\rm SM} \simeq 1,
\label{H_al_lim}
\end{equation}
where $m$ is the mass of the observed Higgs boson, $y_{\rm SM}, y_{\rm BSM}$ are its Yukawa couplings in the SM and the BSM, correspondingly. 
Deviations from the SM predictions in the Higgs sector may be observed in self-interactions of scalar fields, interactions with light quarks and leptons and particle interactions with CP-violation which will be unambiguous evidence of the nonstandard Higgs sector\footnote{
The effects of CP-violating interactions of the SM Higgs boson due to CKM-matrix are so tiny that the observations of them are beyond the experimental possibilities.}
(see \cite{CP-signals}).

Two approaches to the analysis of the Higgs alignment limit are developed:
it can be achieved either through the imposition of symmetries \cite{pil23_21, dev_pilaftsis, pil23_23} or through fine-tuning \cite{pil23_17, pil23_18, pil23_19, pil23_20, 95_haber24, CP-signals}.
The first approach is based on the analysis of
possible symmetries that lead to the SM alignment limit within the 2HDM. It is assumed that at some high mass scale a symmetry exists which ensures naturally an alignment (or universalization) of the Higgs boson interactions with the SM particles
(Natural SM-Higgs Alignment, NHAL). 
It was shown in \cite{dev_pilaftsis} that for the CP-conserving limit three types of symmetries are possible which guarantee NHAL without decoupling 
(simplest symmetry of this sort is $SO(5)$ and the corresponding model is known as the Maximally Symmetric Two Higgs Doublet Model, MS-2HDM). In the case of CP violation, the number of such symmetries increases \cite{pilaftsis_23}. 
At a smaller scale, the symmetry is softly broken, and there will be some deviation from the alignment limit in the low-energy Higgs spectrum as a consequence of the renormalization group effects due to the
hypercharge gauge coupling $g'$ and the third-generation Yukawa couplings \cite{dev_pilaftsis}.
In the second approach, the nature of the alignment limit is not investigated; instead, it is assumed that on the electroweak energy scale, the alignment limit is precisely or approximately fulfilled presenting an inherently fine-tuning example (an effective low-energy approach).
The present analysis is carried out within the framework of the second approach.

In the paper, we consider a supersymmetric (SUSY) extension of the type II 2HDM -- the Minimal Supersymmetric Standard Model (MSSM) \cite{MSSM} -- according to which each SM degree of freedom is associated with a superpartner. As far as no evidence of new particles is observed, we suppose that all SUSY partners are heavy, so the Higgs boson phenomenology at low scale is very similar to that of a type-II 2HDM.  
The current experimental constraints on the masses of SUSY particles and neutral Higgs bosons are $M_{\rm SUSY} > 2.3$ TeV and 
$m_{H,A}>1121$ GeV for $\tan \beta=10$ (and larger for $\tan \beta > 10$) \cite{PDG-24}. 
The mass limit for $m_{H^\pm} < m_{top}$ is $m_{H^\pm} > 155$ GeV, 
in case of $m_{H^\pm} > m_{top}$ it is $m_{H^\pm} > 181$ GeV for $\tan \beta=10$ (and larger for $\tan \beta > 10$) \cite{PDG-24}. 
At the same time, 
the observed deviations in Run 1 and Run 2 at the LHC at an invariant mass of 28 GeV (in a dimuon channel) \cite{28-GeV} or 95 GeV (in channels $\gamma \gamma$, $\tau \tau$, $bb$) \cite{95-GeV} could be a sign of additional Higgs scalars. 

In this work, we consider the Higgs sector of the type II 2HDM and the MSSM with explicit CP-violation and analyze the alignment limit conditions under the assumption that the observed Higgs boson is a neutral scalar with indefinite CP-parity $h_1$ or $h_2$ or $h_3$. In this framework, the possibility of interpretation of the diphoton 95 GeV excess \cite{95_haber24, 95_SSM, 95_mod_ind} (see, however,  \cite{95_not})
 as a possible neutral scalar $h_i$-decay is investigated. 
We propose benchmark scenarios and discuss some 
phenomenological features relevant for future LHC searches.

Note that in the scenario which is considered below the effective field theory at the electroweak scale is the 2HDM, so $M_A$ mass scale is of the order of $m_{top}$ and orders of magnitude below the scale of Higgs superfield mass parameter $\mu$ and the superpartners mass scale $M_S$, the latter are insignificantly different.  Other interesting scenarios have been considered in the literature, for example, the location of the $M_A$ between $m_{top}$ and ($M_S$, $\mu$), when the effective theory above $M_A$ is the 2HDM, below $M_A$ is standard-like, see \cite{lee_wagner, carena_etal}. The effects of new physics including those from explicit CP violation decouple from the light Higgs boson sector, so these interesting cases demonstrate different phenomenology of the heavy Higgs bosons.

\section{Radiative corrections to the Higgs sector of the type-II 2HDM and the MSSM}

The MSSM Higgs boson mass spectrum and mixing of scalars for the effective renormalization group improved potential where the tree-level CP invariance is broken explicitly by Yukawa interactions related to the third generation squarks was first analyzed in \cite{pilaftsis_wagner}. Tree-level couplings of neutral Higgs bosons may be significantly altered by strong mixing with multiple phenomenological consequences. In such a framework
two doublets $\Phi_1$ and $\Phi_2$ of the Higgs sector
with vacuum expectations values (VEVs) $v_1$ and $v_2$
($ v^2=v_1^2+v_2^2=(246 \;\text{GeV})^2$, 
$\tan \beta=v_2/v_1$) form the most general $SU(2)\times U(1)$ renormalizable potential 
\cite{THDM}
\begin{eqnarray}
U & = & 
- \, \mu_1^2 (\Phi_1^\dagger\Phi_1) - \, \mu_2^2 (\Phi_2^\dagger
\Phi_2) - [ \mu_{12}^2 (\Phi_1^\dagger \Phi_2) +h.c.]\nonumber \\
 &&+ \lambda_1
(\Phi_1^\dagger \Phi_1)^2
      +\lambda_2 (\Phi_2^\dagger \Phi_2)^2
+ \lambda_3 (\Phi_1^\dagger \Phi_1)(\Phi_2^\dagger \Phi_2) +
\lambda_4 (\Phi_1^\dagger \Phi_2)(\Phi_2^\dagger \Phi_1) \nonumber \\
&&+ [\lambda_5/2
       (\Phi_1^\dagger \Phi_2)(\Phi_1^\dagger\Phi_2)+ \lambda_6
(\Phi^\dagger_1 \Phi_1)(\Phi^\dagger_1 \Phi_2)+\lambda_7 (\Phi^\dagger_2 \Phi_2)(\Phi^\dagger_1 \Phi_2)+h.c.],
\label{U4}
\end{eqnarray}
where parameters $\mu_{12}, \lambda_{5,6,7}$ can be complex (explicit CP-violation). 
Tree level relations at the SUSY scale define 
quartic couplings as \cite{THDM} 
\begin{equation}
\lambda_{1,2}^{\rm tree} = \frac{g_2^2 + g_1^2}{4}, \qquad
\lambda_{3}^{\rm tree} = \frac{g_2^2 - g_1^2}{4}, \qquad
 \lambda_4^{\rm tree}=-\frac{g_2^2}{2}, \qquad
\lambda_{5,6,7}^{\rm tree}=0.
\end{equation}

At the loop level due to interactions with supersymmetric particles,  parameters $\lambda_i$ acquire threshold corrections 
$\lambda_i = \lambda_i^{\rm tree}+ \Delta \lambda_i^{\rm thr}$. 
Using the renormalization group equations (RGEs), one can evaluate $\lambda_i$ and the corresponding Higgs boson masses
at the electroweak scale ($M_{\rm EW}$) where they can be measured. 
Radiative corrections have a significant impact on the  model predictions. A review of different methods and approaches for radiative correction calculation in the MSSM is presented in \cite{dif_meth_rad_cor} (see also \cite{mod_ph_A}). 
In the case where all non-2HDM states are decoupled, effective field theory (EFT) approach is sufficient. 

An additional type of corrections comes from nonrenormalizable operators of Higgs potential decomposition at the loop level.
Due to self-interactions, the potential $U$, (\ref{U4}), acquires  infinite number of terms of dimension six, eight, and etc., $U_{\rm loop} = U^{(2)}+ U^{(4)}+ U^{(6)}+...$ \cite{cw}\,. 
Within the MSSM, it was found \cite{carena_A_mu} that the decomposition up to dimension four operators is sufficient if
\begin{align}
2 |m_{\rm top} \mu| < M_{\rm SUSY}^2, \qquad |m_{\rm top} A| < M_{\rm SUSY}^2, 
\label{Amu_Ms}
\end{align}
where  $A=A_t=A_b$, $A_{t,b}, \mu$ are soft SUSY breaking parameters, 
$m_{\rm top}$ is the top quark mass.
These model parameters must  also  satisfy an approximate 'heuristic' bound \cite{ew_vac}  according to which the deepest minimum of the effective SUSY potential coincides with the EW minimum
\begin{equation}
\frac{{\rm max}(A_{t,b}, \mu)}{M_{\rm SUSY}} \leq 3,
\end{equation}
so the condition (\ref{Amu_Ms}) is always fulfilled\footnote{
Unstable values of $A_{t,b}, \mu$, however, are also considered in some analyses \cite{unstable_v}\,.
}.

In the general 2HDM without other new fields the corrections to $U^{(4+i)}$  ($i=1,2, ...$) may play an important role to Higgs phenomenology.
The case of nonzero $U^{(6)}$ was considered in \cite{PRD}, where 13 invariant operators of the type $\kappa_i (\Phi_i^\dagger \Phi_j)^3$ were investigated and threshold corrections to parameters $\kappa_i$ were obtained.  
 If the condition (\ref{Amu_Ms}) is not required then one can explain the observed excess in the invariant mass of 28 GeV by taking into account the additional threshold corrections to the dimension six operators \cite{jetp, mod_ph_A}\,.  
 
One must carefully check vacuum stability and perturbative unitarity constraints (see details in \cite{jetp, uzmu_dim6}). 
The conditions for the EW minimum are less constrained, the allowed parameter space is nearly the same as the one considered for the renormalizable Higgs potential (\ref{U4}) \cite{EW_min}.

\begin{figure}[h]
\centerline{\psfig{file=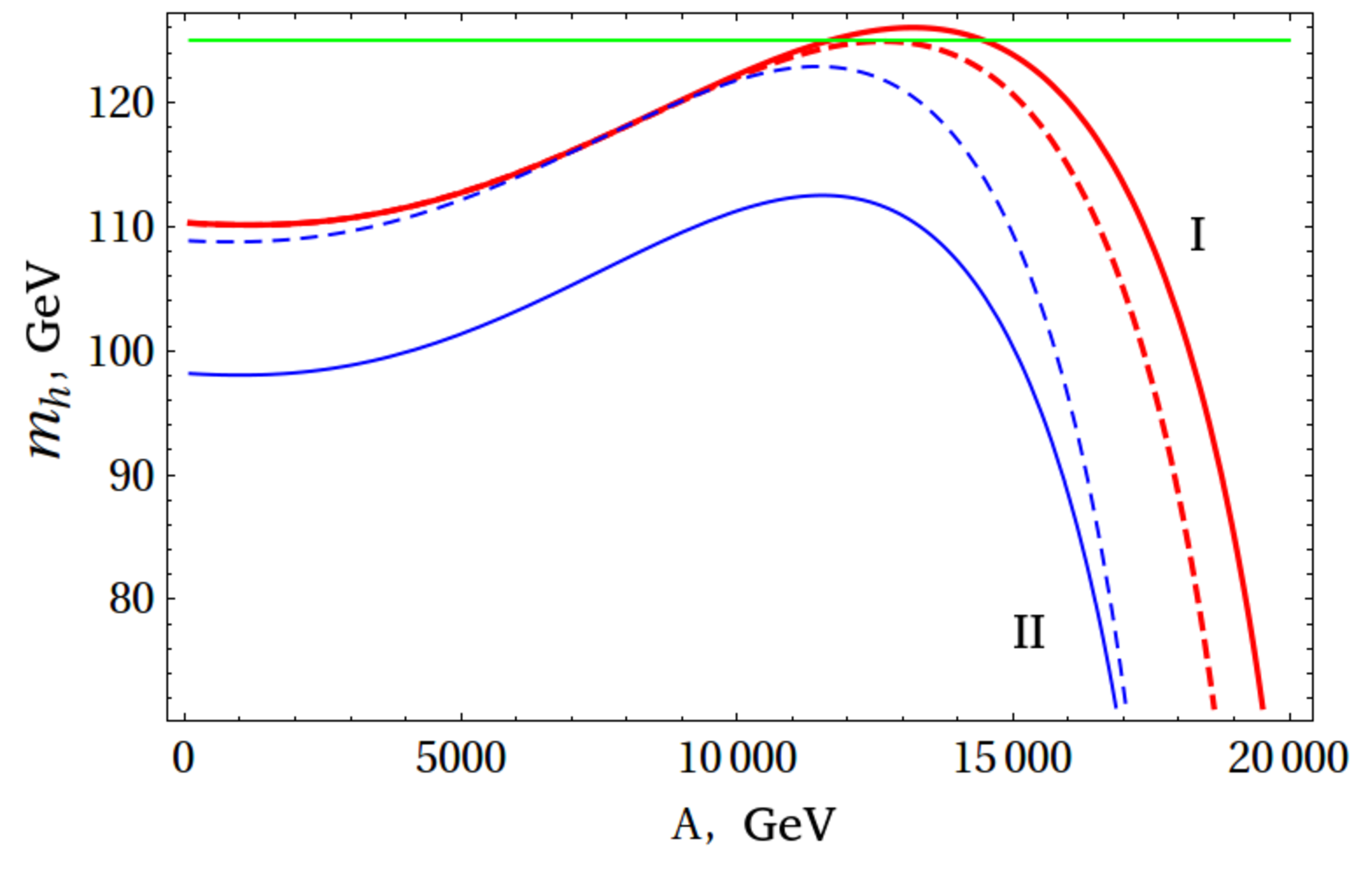,width=8.7cm}}
\vspace*{8pt}
\caption{Higgs mass $m_h$ as a function of $A_{t,b}=A$ with $\kappa_i(M_{\rm SUSY})$ (solid lines) or $\kappa_i(M_{top})$ (dashed lines). 
Here $M_{\rm SUSY}$=3 TeV, $\tan \beta$=5, $m_A=M_{\rm SUSY}$ (red lines) or $m_A$=200 GeV (blue  lines) and (unstable) values of $\mu$=15 TeV, $A_{t,b}/M_{\rm SUSY}>$3, $\mu/M_{\rm SUSY}>$3. 
The horizontal line corresponds to 125 GeV. 
\label{fig:RG-kappa}}
\end{figure}

For correct model predictions, the obtained threshold corrections to $\kappa_i$ must also be evaluated at the EW-scale.  
RG-improvement of scalar potentials in non-renormalizable theories is discussed in \cite{kazakov} where it is assumed that 
divergences are subtracted some way. For potentials similar to our case -- $g \psi^6/6!$ -- there are no RG-analytic expressions and only numeric estimations are possible \cite{kazakov}. 
The loop-level Higgs potential in our case has a more difficult structure, so we evaluate the RG effects to $\kappa_i$ by taking into account the RG-dependence of Yukawa and gauge couplings at the scale $M_{\rm EW}$\footnote{
The explicit representation for $\kappa_i$ and RGEs for $h_{U,D}$, $g_{1,2}$ one can find in \cite{PRD, hh_93}.  
}. 
The dependence of the CP-even Higgs boson mass on the parameter $A$ is presented in Fig.~\ref{fig:RG-kappa} for $m_A$=200 GeV or $m_A=M_{\rm SUSY}$ where $\kappa_i$ are evaluated on the scale $M_{\rm SUSY}$ or $m_{top}$. The predictions differ in both cases, and the most noticeable discrepancy is observed at lower $m_A$. 

\section{Higgs alignment limits in the type II 2HDM and the MSSM with explicit CP-violation}

The Higgs alignment limit (\ref{H_al_lim}) significantly constrains the allowed parameter space of the considered model. In the CP-conserving limit as was discussed in \cite{mod_ph_A, bahl_20}, only the $h$ state can be interpreted as the observed Higgs boson, whereas the SM-like $H$-state is excluded. 
In the type II 2HDM with explicit CPV, we shall assume that the observed Higgs boson is a neutral scalar with an indefinite CP-parity $h_1$ or $h_2$ or $h_3$.
The Higgs states are related as \cite{echaja}
$(h, \, H, \, A)^T = a (h_1, \, h_2, \, h_3)^T$, 
so in the CP-conserving limit the scalar $h_1$ is a light CP-even state $h$, $h_2$ is a heavy CP-even scalar $H$, $h_3$ is a CP-odd state $A$. Thus $m_{h_1}(\varphi=0) \leq m_{h_2}(\varphi=0)$ where
$\varphi= \arg (A \mu)$ is a CP-violating phase.  
The matrix $a$ can be presented as $a_{ij}=a_{ij}^{'}/n_j$, $n_j=k_j \sqrt{a_{1j}^{'2}+a_{2j}^{'2}+a_{3j}^{'2}}$ ($k_j=\pm 1$) \cite{echaja, cern_yellow}
\begin{eqnarray}
a_{11}^{'} &=& [(m_H^2-m_{h_1}^2)(m_A^2-m_{h_1}^2)-c_2^2], \quad
a_{12}^{'} = -c_1 c_2, \quad
a_{13}^{'} = -c_1(m_H^2-m_{h_3}^2), \nonumber \\
a_{21}^{'} &=& c_1 c_2, \quad
a_{22}^{'} = -[(m_h^2-m_{h_2}^2)(m_A^2-m_{h_2}^2)-c_1^2], \quad
a_{23}^{'} = -c_2(m_h^2-m_{h_3}^2), \nonumber \\
a_{31}^{'} &=& -c_1(m_H^2-m_{h_1}^2), \,
a_{32}^{'} = c_2(m_h^2-m_{h_2}^2),  \,
a_{33}^{'} = (m_h^2-m_{h_3}^2)(m_H^2-m_{h_3}^2),
\end{eqnarray}
where in the case of Higgs potential decomposition up to dim-six operators \cite{PRD} 
\begin{eqnarray}
c_1 &=& v^2(-1/2 \cdot {\rm Im}\lambda_5 c_{\alpha+\beta}+{\rm Im}\lambda_6 s_\alpha c_\beta-{\rm Im}\lambda_7 c_\alpha s_\beta) 
+ \frac{v^4}{4} [ -c_{\alpha+\beta} s_{2 \beta} (3 {\rm Im} \kappa_7 \nonumber  \\
&&+{\rm Im} \kappa_{11}+{\rm Im} \kappa_{13}) + 4(s_\alpha c_\beta^3 {\rm Im}\kappa_8-c_\alpha s_\beta^3 {\rm Im} \kappa_{12}) \nonumber \\ 
&& + 2[ s_\beta^2 (-3 c_\alpha c_\beta+s_\alpha s_\beta){\rm Im}\kappa_{10}
-c_\beta^2 (c_\alpha c_\beta-3 s_\alpha s_\beta){\rm Im}\kappa_{9}]\}, 
\\
c_2 &=& -\frac{v^2}{2} \{ {\rm Im}\lambda_5 s_{\alpha+\beta}+2 ({\rm Im} \lambda_6 c_\beta c_\alpha+{\rm Im}\lambda_7 s_\beta s_\alpha) 
+ v^2[2{\rm Im}\kappa_8 c_\beta^3 c_\alpha \nonumber \\
&& +{\rm Im}\kappa_9 c_\beta^2 (s_{\alpha+\beta}+2c_\alpha s_\beta) + {\rm Im}\kappa_{10} s_\beta^2 (s_{\alpha+\beta}+2c_\beta s_\alpha)
+ 2 {\rm Im}\kappa_{12} s_\beta^3 s_\alpha \nonumber \\
&& +\frac{1}{2}(3{\rm Im}\kappa_{7}+{\rm Im}\kappa_{11}+{\rm Im}\kappa_{13})s_{2 \beta} s_{\alpha+\beta}] \} ,
\end{eqnarray} 
$\beta, \alpha$ are mixing angles in the Higgs sector
[$\alpha \in (- \pi/2,0]$, $\beta \in (0,\pi/2)$]
\begin{equation}
\tan 2 \alpha = \frac{2 \Delta{\cal M}_{12}^2 - (m_Z^2+m_A^2)s_{2\beta}}{(m_Z^2-m_A^2)c_{2\beta} +\Delta {\cal M}_{11}^2- \Delta {\cal M}_{22}^2}, 
\end{equation}
$\Delta {\cal M}_{ij}^2$ are radiative corrections to the CP-even mass matrix \cite{PRD}\,. 

In order to find out the $h_i$-alignment limit conditions ($i=1,2,3$), we analyze the following forms of Higgs interactions with
up and down SM fermions and gauge bosons
\cite{dub_sem, echaja}
\begin{eqnarray}
g(h_i uu) &=& (s_\alpha a_{2i} +c_\alpha a_{1i} - ic_\beta a_{3i}\gamma_5)/s_\beta, \nonumber  \\
g(h_i dd) &=& (c_\alpha a_{2i} -s_\alpha a_{1i} - is_\beta a_{3i}\gamma_5)/c_\beta, \nonumber \\
g(h_i VV) &=& c_{\beta-\alpha} a_{2i} +s_{\beta-\alpha} a_{1i}.
\label{g_huu}
\end{eqnarray}
Then the Higgs alignment limit conditions can be presented as
\begin{subequations} 
\label{h12_al} 
\begin{eqnarray}
h_1:&& \,
{\rm (I}) \, \beta - \alpha \simeq \pi/2, \, c_1 \simeq 0;  \quad
{\rm (II}) \, \tan(\beta-\alpha) \simeq -c_2/c_1, \, m_{h_1} \simeq m_H ; \label{h1_al} \\
h_2:&& \, 
{\rm (I}) \, \alpha \simeq 0,  \, \beta \simeq 0, \, c_2 \simeq 0, \quad
{\rm (II}) \, \tan(\beta-\alpha) \simeq -c_2/c_1, \, m_{h_2} \simeq m_h;
\label{h2_al}\\
h_3:&& \, {\rm (I}) \, \alpha \simeq 0,  \quad \beta \simeq 0, \, m_{h_3} \simeq m_H;  \quad
{\rm (II}) 
\, \beta - \alpha \simeq \pi/2, \quad m_{h_3} \simeq m_h,
\label{h3_al}
\end{eqnarray}
\end{subequations}
where for each $h_i$-alignment two different sets of conditions are possible. 
The $h_1^{\rm I}$-alignment limit\footnote{
This case was investigated in \cite{uzmu_CP}.
}, (\ref{h1_al}), resembles the one in the model with CPC added by relation $c_1 \simeq 0$
which fixes the CP-violating phase 
as solutions of equation
$
a c_\varphi^2 + b c_\varphi + c \simeq 0
$,
where
\begin{eqnarray*}
 a &=& -3 v^2 c_{\alpha+\beta} s_\beta |\kappa_7|,\\
 b &=& - c_{\alpha+\beta} |\lambda_5| +v^2 [s_\beta^2 (-3 c_\alpha  c_\beta+s_\alpha s_\beta) |\kappa_{10}| - c_\beta^2 (c_\alpha c_\beta-3 s_\alpha s_\beta) |\kappa_9| ],\\
 c &=&  s_\alpha c_\beta |\lambda_6| -  c_\alpha s_\beta |\lambda_7| + \frac{v^2}{4} \left[ 3 c_{\alpha+\beta} s_\beta |\kappa_7|+|\kappa_{11}|+|\kappa_{13}|+4 (s_\alpha c_\beta^3 |\kappa_8| - c_\alpha s_\beta^3 |\kappa_{12}|) \right] .
\end{eqnarray*}
In the limit  $\kappa_i$=0 the phase is defined by
\begin{equation}
\cos \varphi = \frac{|\lambda_6| s_\alpha c_\beta -|\lambda_7| c_\alpha s_\beta}{|\lambda_5| c_{\alpha+\beta}}.
\label{varphi}
\end{equation}

Alignments  $h_2^{\rm I}$ and $h_3^{\rm I}$ are valid for $\cos(\beta - \alpha)=1$ and as far as the only point where $\alpha$ and $\beta$ are close to each other is 0, we end up with conditions $\alpha \simeq 0$,   $\beta \simeq 0$.  
The choice of $\tan \beta \simeq 0$ is not relevant for phenomenology as it leads to a massless $b$ quark, so we rule out the $h_2^{\rm I}$ and $h_3^{\rm I}$ alignments. 

The last $h_1^{\rm II}$, $h_2^{\rm II}$ and $h_3^{\rm II}$ alignments 
can be realized with Higgs boson masses of the EW-scale $m_{h_i} \sim M_{\rm EW}$. 
Taking into account the current experimental bounds for searching Higgs neutral scalars \cite{PDG-24}, the alignment limits with $m_{h_i} \sim M_{\rm EW}$ for $\tan \beta \geq$10 are excluded. For the case of $\tan \beta \leq$10 the situation is more unambiguous as far as we do not know the experimental constraints on the neutral Higgs bosons in this region. 
Numerical estimations performed for fixed parameters
\begin{equation}
\tan \beta =\{2, \, 7\}, \qquad
M_{\rm SUSY} = \{ 2.5, \, 5\} \text{ TeV}, \qquad
m_A = \{ 96, \, 125 \} \text{ GeV}
\end{equation}
and varied parameters $(|A|, |\mu|) \in [-3 \div 3] \times M_{\rm SUSY}$ and $\varphi \in (0, 2 \pi)$ 
reveal no acceptable parameter region satisfying the corresponding alignment limit conditions.
It is easily to provide $m_{h_1}$=95 GeV, $m_{h_2}$=125 GeV but the alignment conditions for mass relations are never satisfied. 
Thus only the Higgs alignment limit $h^{\rm I}_1$ is realized in framework of the MSSM.

The alignment limit conditions (\ref{h1_al}) allow to predict CP-violating interactions of Higgs bosons with SM particles in general form. Analyzing (\ref{g_huu}) we can notice that CPV signals may be observed only in  interactions of $h_3$ with SM fermions if the value $a_{33}$ is large enough. 

\section{Benchmark scenarios}
To analyze MSSM predictions for the parameter space 
\begin{equation}
\tan \beta, \qquad 
M_{\rm SUSY}, \qquad  m_{H^{\pm}}, \qquad 
|A_{t,b}|=|A|, \qquad |\mu|, \qquad  \varphi,
\end{equation}
satisfying the alignment limit conditions $h_1^{\rm I}$, we scan it
and obtain model regimes (benchmark scenarios, BS). 
Numerical analysis is performed in framework of EFT-approach (see \cite{echaja, mod_ph_A, PRD} and Refs therein) in the approximation of  degenerate squark masses of third generation\footnote{
Note that the most accurate and complete computations of Higgs boson masses within the EFT approach can be implemented in framework of the CPsuperH3.0 \cite{carena_etal, CPsH}.
However, assuming the theoretical uncertainty of $m_{h_1}=125 \pm 2$ GeV we restrict ourself by the approximation mentioned above.
}. 
The CP-violating phase $\varphi$ is an input parameter defined by (\ref{varphi}).
We fix
\begin{equation}
\tan \beta =\{2, \, 5, \, 10, \, 20 \}, \qquad
M_{\rm SUSY} = \{ 2.5, \, 5, \, 10 \} \text{ TeV}, \qquad
m_{H^{\pm}} = \{ 300, \, 3000 \} \text{ GeV}
\end{equation}
and varied parameters $(|A|, |\mu|) \in [-3 \div 3] \times M_{\rm SUSY}$  in such a way that the alignment limit $h_1^{\rm I}$ ($m_{h_1}$=125 GeV, $\beta - \alpha \simeq \pi/2 $, $c_1 \simeq 0$)
is satisfied. 

No allowed region for $\tan \beta <$ 5 is found.
The large mass of the charged Higgs boson $m_{H^\pm} \geq M_S$ is more preferable. In this case all additional Higgs bosons decouple and are nearly degenerated. 
Obtained benchmark scenarios are presented in Table~\ref{table}. 
Note that BS3 is close to the benchmark scenario {\it CPX4LHC} proposed in \cite{carena_etal} in the case $\mu=2 M_{\rm SUSY}$, however, the obtained alignment limit conditions, (\ref{h1_al}), rule out the mass $m_{H^\pm}$ up to 4 TeV.

Within these scenarios the dependence of $a_{33}$ can be analyzed. It turns out that prediction for $a_{33}$ is insensitive to $m_{H^\pm}$ and almost constant. The obtained values are presented in Table~\ref{table}. We can conclude straightforwardly [see (\ref{g_huu})] that rather significant CPV interactions of $h_3$ with up-fermions take place in  scenarios BS2 and BS4. 

\begin{table}[h!]
\caption{Benchmark scenarios satisfying the alignment limit conditions $h_{1}^{\rm I}$ ($\beta - \alpha \simeq \pi/2$, $c_1 \simeq 0$) with the accuracy $m_{h_1}=125 \pm 2$ GeV and $|c_1|$ less than 0.1 or 0.01.}
{\begin{tabular}{cccccccc} 
\hline
\hline
BS & $\tan \beta$ & $M_{\rm SUSY}$ & $A$ & $\mu$ & $m_{H^\pm}$ & $|c_1|$ & $a_{33}$ \\
& & (TeV) & (TeV) & (TeV) & (TeV) & (accuracy) & (prediction) \\ 
\hline
BS1 & 5 & 2.5 & 5.5 & 1 & varied ($\geq 1$) &  $\leq 0.01$ & 0.008\\
BS2 & 10 & 2.5 & 5.5 & 3 & varied ($\geq 3$) & $\leq 0.01$ & 0.288 \\
BS3 & 10 & 5 & 10 & varied ($1-10$) & varied & $\leq 0.01$ & 0.026 \\
	&	&	&	& 1 & $\geq 1$\\
		&	&	&	& 10 & $\geq 4$\\
BS4 & 20 & 10 & 28 & 12 & varied ($\geq 3$) & & 0.336 \\
	&	&	&		& & 3 & $\leq 0.1$\\
	&	&	&		& & 10 & $\leq 0.01$\\
\hline
\hline
\end{tabular} \label{table}}
\end{table}

\section{Conclusion}
We have considered the two-Higgs doublet sector with explicit CP-violation where the effective Coleman-Weinberg type and RG-improved Higgs potential are analyzed within the framework of decomposition up to dimension-six operators.
We have also implemented the threshold corrections to the MSSM RG running, induced by dimension-six operators. 
We have found that the mass shifts of a light neutral Higgs scalar $h$ are negligible for stable values of $A_{t,b}/M_{\rm SUSY} \leq $3, $\mu/M_{\rm SUSY}\leq $3 and are about 10 GeV at $m_A \sim M_{\rm EW}$ or negligible at $m_A \sim M_{\rm SUSY}$ for unstable values $A_{t,b}/M_{\rm SUSY}>3$, $\mu/M_{\rm SUSY}>3$. 

For the general type II 2HDM, the Higgs alignment limit conditions have been obtained which are valid within the dimension-six decomposition of the effective Higgs potential for the neutral Higgs bosons with indefinite CP-parity $h_1, h_2$ or $h_3$.  
Numerical investigations within the MSSM reveal that the only $h_1^{\rm I}$-alignment limit takes place. 
For any $\tan \beta$ parameter, additional Higgs bosons decouple and are
heavier than 1 TeV. 
No possibility for a neutral Higgs scalar with a mass of 95 GeV remains.
For investigating the main phenomenological features relevant for future collider searches, four benchmark scenarios have been proposed. 
Two of them predict distinguishable CP-violating interactions of the Higgs boson $h_3$ with up-fermions. 

Note that the above analysis refers to the generic basis for the type II 2HDM potential. CP-violating flavor-aligned 2HDM was also analyzed in \cite{95_haber24} where the Higgs alignment limit conditions were obtained for the $h_1$ state in the Higgs basis. Basis-independent methods developed for the 2HDM in \cite{davidson_haber} allow in principle to construct
explicit representations for the couplings and mixing angles in the mass eigenstate basis in terms of 2HDM invariants (quantities that are scalar under arbitrary
unitary transformations among the two Higgs fields in the Lagrangian) also for the extended Higgs potential, which is beyond the scope of this work.

\section*{Acknowledgments}
The work of E.F. was supported by the Theoretical Physics and Mathematics Advancement Foundation "BASIS".


\begin{thebibliography}{0}   

\bibitem{obs-125}
ATLAS Collab. (G. Aad \emph{et al.}), 
{\it Phys. Lett. B} \textbf{716} (2012) 1,
arXiv:1207.7214 [hep-ex];\\
CMS Collab. (S. Chatrchyan \emph{et al.}),
{\it Phys. Lett. B} \textbf{716} (2012) 30,
arXiv:1207.7235 [hep-ex].

\bibitem{BEH-mech}
F.~Englert and R.~Brout, 
{\it Phys. Rev. Lett.} \textbf{13} (1964) 321;
P.~W.~Higgs, 
{\it Phys. Lett.} \textbf{12} (1964) 132.

\bibitem{H_presic}
ATLAS Collab. (G.  Aad \emph{et al.}), 
{\it Nature} \textbf{607}, No. 7917 (2022) 52,
arXiv:2207.00092 [hep-ex];
%
CMS Collab. (A. Tumasyan \emph{et al.}), 
{\it Nature} \textbf{607}, No. 7917 (2022) 60, 
arXiv:2207.00043 [hep-ex].

\bibitem{PDG-24}
Particle Data Group (S. Navas \emph{et al.}), 
{\it Phys. Rev. D} \textbf{110} (2024) 030001.

\bibitem{THDM}
J. F. Gunion, H. E. Haber et al.,  
{\it The Higgs Hunter's Guide},
Addison-Wesley, 1990.

\bibitem{echaja}
E.~Akhmetzyanova, M.~Dolgopolov and M.~Dubinin, 
{\it Phys. Rev. D} \textbf{71} (2005) 075008, 
arXiv:hep-ph/0405264; 
{\it Phys. Part. Nucl.} \textbf{37} (2006)  677.

\bibitem{align}
M.~Carena \emph{et al.}, {\it Phys. Rev. D.} \textbf{91} (2015) 3,  035003,
arXiv:1410.4969 [hep-ph];
%
D.~Asner \emph{et al.}, arXiv:1310.0763 [hep-ph].

\bibitem{CP-signals}
H.~E. Haber, V. Keus and R. Santos,
{\it Phys. Rev. D} \textbf{106}, No 9 (2022) 095038, 
arXiv:2206.09643 [hep-ph]. 

\bibitem{pil23_21}
A. Pilaftsis, 
{\it Phys. Rev. D} {\bf 93} (2016) 075012, 
arXiv:1602.02017 [hep-ph].

\bibitem{dev_pilaftsis}
P.~S.~B. Dev and A. Pilaftsis, 
{\it JHEP} {\bf 12} (2014) 024; {\it JHEP} {\bf 11} (2015) 147 (erratum), 
arXiv:1408.3405 [hep-ph].

\bibitem{pil23_23}
N. Darvishi and A. Pilaftsis, 
{\it Phys. Rev. D} {\bf 101} (2020) 095008,
arXiv:1912.00887 [hep-ph].

\bibitem{pil23_17}
P.~H. Chankowski, T. Farris, B. Grzadkowski, J.~F. Gunion, J. Kalinowski and M. Krawczyk, 
{\it Phys. Lett. B} {\bf 496} (2000) 195, 
arXiv:hep-ph/0009271.

\bibitem{pil23_18}
J.~F. Gunion and H.~E. Haber, 
{\it Phys. Rev. D} {\bf 67} (2003) 075019, 
arXiv:hep-ph/0207010.

\bibitem{pil23_19}
I.~F. Ginzburg and M. Krawczyk, 
{\it Phys. Rev. D} {\bf 72} (2005) 115013,
arXiv:hep-ph/0408011.

\bibitem{pil23_20}
M. Carena, I. Low, N.~R. Shah and C.~E.~M. Wagner, 
{\it JHEP} {\bf 04} (2014) 015,
arXiv:1310.2248 [hep-ph].

\bibitem{95_haber24}
S.~Banik, G.~Coloretti, A.~Crivellin and H.~E.~Haber,
arXiv:2412.00523v2 [hep-ph]. 

\bibitem{pilaftsis_23}
N. Darvishi, A. Pilaftsis, and J.-H. Yu,
{\it JHEP} {\bf 05} (2024) 233,
arXiv:2312.00882 [hep-ph].

\bibitem{MSSM}
H.~Haber and G.~Kane,
{\it Phys. Rept.} \textbf{117} (1985) 75.

\bibitem{28-GeV}
CMS  Collab. (A.~M.~Sirunyan \emph{et al.}),
{\it JHEP}
\textbf{11} (2018) 161,
arXiv:1808.01890 [hep-ex]; 
%
{\it JHEP } \textbf{11} (2017) 010,
arXiv:1707.07283 [hep-ex].

\bibitem{95-GeV}
CMS Collab. (A. Hayrapetyan \emph{et al.}),
CMS-PAS-HIG-20-002, 
arXiv:2405.18149 [hep-ex];
%
ATLAS Collab., 
ATLAS-CONF-2023-035;
%
CMS Collab. (A. Tumasyan \emph{et al.}),
{\it JHEP} \textbf{07} (2023) 073,
arXiv:2208.02717 [hep-ex];
%
LEP Working Group for Higgs boson searches,
ALEPH, DELPHI, L3, OPAL Collab. (R. Barate \emph{et al.}), 
Phys. Lett. B \textbf{565} (2003) 61, 
arXiv:hep-ex/0306033.

\bibitem{95_SSM}
C.-X.~Liu  \emph{et al.}, 
{\it Phys. Rev. D} \textbf{109}, No 5, 056001 (2024).
arXiv:2402.00727 [hep-ph]. 

\bibitem{95_mod_ind}
T. Mondal, S. Moretti and P. Sanyal, 
arXiv:2412.00474 [hep-ph].

\bibitem{95_not}
P. Janot,
{\it JHEP} \textbf{10} (2024) 223, 
arXiv:2407.10948v2 [hep-ph].

\bibitem{lee_wagner}
G. Lee and C.~E.~M. Wagner, 
{\it Phys.Rev. D} {\bf 92} (2015) 075032,
arXiv:1508.00576 [hep-ph]. 

\bibitem{carena_etal}
M. Carena, J. Ellis, J.~S. Lee, A. Pilaftsis and C.~E.~M. Wagner,
{\it JHEP} {\bf 02} (2016) 123,
arXiv:1512.00437 [hep-ph].

\bibitem{pilaftsis_wagner}
A. Pilaftsis and C. Wagner, 
{\it Nucl.Phys. B} {\bf 553} (1999) 3,
arXiv:hep-ph/9902371.


\bibitem{dif_meth_rad_cor}
P. Draper and H. Rzehak, 
{\it Phys. Rept.} \textbf{619} (2016) 1, 
arXiv:1601.01890 [hep-ph].

\bibitem{mod_ph_A}
M.~N. Dubinin and E.~Yu. Petrova, 
{\it Int. J. Mod. Phys. A} \textbf{33} (2018) 1850150,
arXiv:1709.10301 [hep-ph].

\bibitem{cw}
S. Coleman and E. Weinberg, {\it Phys. Rev. D} \textbf{7} 
(1973) 1888.

\bibitem{carena_A_mu}
M. Carena, J. R. Espinosa, M. Quiros and C. E. M. Wagner, 
{\it Phys. Lett. B} \textbf{355} (1995) 209,
arXiv:hep-ph/9504316.

\bibitem{ew_vac}
W.~G.~Hollik, G.~Weiglein and J.~Wittbrodt, 
{\it JHEP} \textbf{03} (2019) 109,
arXiv:1812.04644 [hep-ph].

\bibitem{unstable_v}
P. Bechtle \emph{et al.}, 
{\it Eur. Phys. J. C} 77, No 2 (2017) 67, 
arXiv:1608.00638 [hep-ph].

\bibitem{PRD}
M.~N. Dubinin and E.~Yu. Petrova, 
Phys. Rev. D. \textbf{95} (2017) 055021,
arXiv:1612.03655 [hep-ph].

\bibitem{jetp}
M.~N. Dubinin and E.~Yu. Fedotova,
{\it JETP} \textbf{131}, No. 6 (2020) 917.

\bibitem{uzmu_dim6}
M.~N. Dubinin and E.~Yu. Fedotova,
{\it Memoirs of the Faculty of Physics}, 
No 4 (2022) 2241503, 
http://uzmu.phys.msu.ru/abstract/2022/4/2241503/.

\bibitem{EW_min}
M.~N. Dubinin and E.~Yu. Fedotova, 
{\it EPJ WoC} \textbf{158} (2017) 02005.

\bibitem{kazakov}
D.~I. Kazakov, R.~M. Iakhibbaev and D.~M. Tolkachev, 
{\it JHEP} \textbf{04} (2023) 128, 
arXiv:2209.08019 [hep-th].

\bibitem{hh_93}
H.~E.~Haber and R. Hempfing, 
{\it Phys. Rev. D} \textbf{48} (1993) 4280,
arXiv:hep-ph/9307201; 
%
{\it Phys. Rev. Lett.} \textbf{66} (1991) 1815.  

\bibitem{bahl_20}
E.~Bagnaschi \emph{et al.},
{\it Eur. Phys. J. C} \textbf{79}, No 7 (2019) 617, 
arXiv:1808.07542 [hep-ph].

\bibitem{cern_yellow}
E. Accomando \emph{et al.},
arXiv:hep-ph/0608079. 

\bibitem{uzmu_CP}
M.~N. Dubinin and E.~Yu. Fedotova,
{\it Memoirs of the Faculty of Physics}, 
No 4 (2023) 2341506,
http://uzmu.phys.msu.ru/abstract/2023/4/2341506/.

\bibitem{dub_sem}
M.~N.Dubinin and A.~V. Semenov, 
{\it Eur. Phys. J. C} \textbf{28} (2003) 223, 
arXiv:hep-ph/0206205v3.


\bibitem{CPsH}
J.~S. Lee  \emph{et al.}, 
{\it Comput. Phys. Commun.} \textbf{156} (2004) 283, 
arXiv:hep-ph/0307377. 

\bibitem{davidson_haber}
S. Davidson and H. Haber, {\it Phys.Rev. D} {\bf 72} (2005) 035004,
arXiv:hep-ph/0504050.

\end{thebibliography}
\end{document}